\documentclass[aps,preprint, superscriptaddress]{revtex4}
\begin{document}
\renewcommand{\topfraction}{1}

\renewcommand{\=}{\!\!=\!\!}

\title{D-term inflation without cosmic strings}
\author{J. Urrestilla}
\affiliation{Department of Physics and Astronomy, University of Sussex, Brighton, BN1 9QJ, UK}
\affiliation{Department of Theoretical Physics, UPV-EHU,
Bilbao, Spain} 
\author{A. Ach\'{u}carro} 
\affiliation{Lorentz Institute of
Theoretical Physics, University of Leiden, 2333 RA Leiden, The
Netherlands} 
\affiliation{Department of Theoretical Physics, UPV-EHU,
Bilbao, Spain} 
\author{A.C. Davis} 
\affiliation{DAMTP, Centre for
Mathematical Sciences, University of Cambridge, CB3 0WA, UK}
\date{\today}
\begin{abstract}
We present a superstring-inspired version of D-term inflation which
does not lead to cosmic string formation and appears to satisfy the
current CMB constraints. 
It differs from minimal
D-term inflation by a second pair of charged superfields which 
makes the strings non-topological (semilocal).  The strings are also
BPS, so the scenario is expected to survive
supergravity corrections.
The second pair of charged superfields arises
naturally 
in several brane
and conifold scenarios, but its effect on cosmic string formation had not been noticed so far. 
\end{abstract}
\maketitle

\def\rd{{\rm{d}}}
\def\d{\partial}
\def\half{{1 \over 2}}
\def\quarter{{1 \over 4}}
\def\thetabar{\bar{\theta}}
\def\sech{{\rm sech}}
\def\tanh{{\rm tanh}}
\def\simleq{\; \raise0.3ex\hbox{$<$\kern-0.75em
      \raise-1.1ex\hbox{$\sim$}}\; }
\def\simgeq{\; \raise0.3ex\hbox{$>$\kern-0.75em
      \raise-1.1ex\hbox{$\sim$}}\; }  
\def\L{{\mathcal L}}
\def\t{\tilde}      
\def\bea{\begin{eqnarray}}
\def\eea{\end{eqnarray}}
\def\be{\begin{eqnarray}}
\def\ee{\end{eqnarray}}

The WMAP \cite{wmap} data have put in sharp focus the question of
identifying the microscopic origin of the inflaton, the field or
fields that fuel inflation.  D-term inflation in supersymmetric (SUSY) gauge
theories is a promising hybrid inflation scenario which
seems to survive supergravity corrections \cite{S95,DSS94}
(see \cite{KL03} for a recent discussion). Unfortunately it suffers
 from a severe drawback: at the
end of inflation cosmic strings form giving a contribution to the
primordial density perturbations comparable to that of
inflation, in  contradiction with the CMB data \cite{LR99}.

While this problem may be circumvented in various ways (for instance,
by invoking a curvaton \cite{EKM03}), much of the beautiful simplicity
and predictive power of the original minimal D-term inflation model is
lost in these modifications. It would be desirable to have a model
which has all the good features of the model of \cite{S95,DSS94}
but without the production of cosmic strings. The purpose of this
letter is to present such a model.

The idea is extremely simple. The minimal D-term model is a
supersymmetric Abelian gauge model where the inflaton is a neutral
scalar field that couples to two charged scalar fields with opposite
$U(1)$ charges. The model has a Fayet-Iliopoulos (FI) D-term. At the
end of inflation, the charged fields acquire non-zero v.e.v.s, the
gauge field is Higgsed and Nielsen-Olesen (NO) strings form. To avoid
string formation, we include a second identical pair of charged fields
with the same couplings to the inflaton and to the gauge field. At the
end of inflation the charged scalars acquire v.e.v.s and cause a Higgs
mechanism, as before, but now the vacuum manifold is simply connected
and the strings are non-topological. They are semilocal (SL) strings
\cite{VA91,H92}, whose cosmological formation rates have been studied
in \cite{ABL99} and whose number density falls to zero when the
couplings are in the {\it Bogomol'nyi} limit - the region of interest
here.

The addition of the second pair of charged fields is motivated by a
possible interpretation as four-dimensional effective actions of type
II superstrings compactified on Calabi-Yau (CY) manifolds. These are expected
to have $N\=2$ SUSY  in four dimensions. Each pair of charged $N\=1$
superfields forms an $N\=2$ hypermultiplet, and the two $N\=2$
hypermultiplets have opposite $U(1)$ charges.  The neutral $N\=1$
superfield (the inflaton) and the $U(1)$ gauge multiplet combine into an
$N\=2$ vector multiplet.
 
D-term inflation has been studied in heterotic string
compactifications where the FI terms arise from an anomalous $U(1)$
\cite{CM89}. Instead, the model presented here seems to be
related to type II string compactifications on CY spaces near singular
points (see also \cite{H03}). The hypermultiplets are typically
associated with string or brane states wrapping around cycles in the
internal space whose size goes to zero at the singularity, making the
states massless. The FI terms appear when the
singularity is resolved by replacing the singular point with a small
``sphere'' (whose size determines the FI parameter).
FI
terms may also arise in brane-antibrane backgrounds \cite{DKP},
and therefore in brane inflation \cite{DT99}. 
Thus, the  analysis of these models
gives a potential window into string theory \cite{H03,DKP,DV}.

Specific examples have been studied, e.g., in connection with
topology change transitions in CY spaces \cite{S,GMS}, phases
of SUSY  Abelian theories \cite{W93} and confinement
\cite{GMV96} in string theory, where the presence of the second,
oppositely charged hypermultiplet was forced by the homology of the
vanishing cycles or chains \footnote{ The example in
\cite{GMS,GMV96} is type IIB strings compactified on a CY conifold
which contains 16 singular points subject to one homology relation,
leading to 16 hypermultiplets charged under 15 $U(1)$s in the low
energy theory.}.

If we ignore gravity, the model presented here has  $N\=2$
SUSY. Coupling to gravity breaks $N\=2$ to (local) $N\=1$ due to
the FI terms
(see
\cite{DKP,AU04,K01} and references therein).  A recent study 
\cite{DKP} shows that the NO strings produced after inflation in the
one-hypermultiplet model are BPS strings (that is, they break half of
the $N\=1$ {\it local} SUSY, like in the global case \cite{DDT}). 
For SUSY to be (partially) unbroken, the transformation of the
gravitino (and all other fermions) must be zero. The holonomy of the
local SUSY parameter going around the string receives two
contributions, one from the deficit angle in the metric transverse to
the string, the other from an Aharonov-Bohm interaction with the
$U(1)$ magnetic field of the string. Cancellation of these two
contributions allows for the existence of killing spinors and unbroken
SUSY \cite{BBS95}, provided the gravitino is charged under the $U(1)$,
which is therefore an R-symmetry. We will show that in our model the
(embedded) NO strings are also BPS but are not expected to form after
inflation.

\section{The model}

Minimal D-term inflation (see \cite{DSS94} for instance)
requires two chiral superfields $\phi_+, \ \phi_-$, with opposite
$U(1)$ charges $\pm g$, a neutral chiral superfield $S$ and a $U(1)$
gauge multiplet. We add another pair of charged chiral superfields
$\tilde\phi_+, \ \tilde\phi_-$ \cite{ADPU02}. In the absence of gravity, the
model has $N\=2$ SUSY, so each pair of chiral superfields
assemble into a non-chiral $N\=2$ hypermultiplet; and the neutral
superfield and the gauge multiplet into an $N\=2$ vector
multiplet. Each hypermultiplet transforms under the $SU(2)_R$ symmetry
between the two supercharges. It is always possible to add an $N\=2$ FI
$\vec{k}\cdot\vec{P}$ term (a {\it P-term}) where $\vec{P}$ are the
$SU(2)_R$ triplet of auxiliary fields belonging to the $U(1)$ Abelian
vector multiplet.  The choice $\vec{k} \propto (0,0,1)$
leads to $N\=1$ D-terms, while $\vec{k} \propto (1,0,0) $ leads to
F-terms. Here we take $\vec{k}=(0,0,g\omega^2/2)$.

In global supersymmetry this choice can be made without loss of
 generality, and shows that D-term and F-term models are equivalent,
 by $SU(2)_R$ rotations, and part of a larger class of so-called
 P-term models \cite{KL03}.  Coupling to $N\=1$ supergravity breaks
 this equivalence \cite{K01,KL03}. Loosely speaking, if the direction
 of supergravity is ``aligned'' with the FI terms, one obtains a
 D-term model. If it is misaligned, F-term or P-term inflation
 results. The higher dimensional origin of the FI terms leads us to
 suppose that the breaking of supergravity is triggered by the FI
 terms themselves and therefore aligned with them \cite{AU04}, so the
 four-dimensional model is a D-term model.

The full matter Lagrangian and supersymmetry transformations can
be found in \cite{ADPU02,ADPU03}. After eliminating auxiliary fields,
the bosonic part reads: 
\bea
 \L&=&\half |D_\mu \phi_+|^2+\half |D_\mu
\phi_-|^2 +\half |D_\mu \tilde\phi_+|^2+\half |D_\mu \tilde\phi_-|^2\nonumber\\
& &
+\half |\d_\mu S|^2 -\quarter F^{\mu \nu}F_{\mu \nu}-V
\label{lagrangian}
\eea
with the tree-level scalar potential $V$ given by 
\bea
\lefteqn{V=\frac{g^2}{2}\left\{\quarter\left[|\phi_+|^2+|\tilde\phi_+|^2-|\phi_-|^2-|\tilde\phi_-|^2-\omega^2\right]^2+\right.}\\
& &\left.
\left|\phi_+\phi_--\tilde\phi_-\tilde\phi_+\right|^2+S^2\left[|\phi_+|^2+|\phi_-|^2+|\tilde\phi_+|^2+|\tilde\phi_-|^2\right]\right\}\nonumber
\label{pot}
\eea 
where $A_\mu$ is the $U(1)$ gauge field, $D_\mu=\partial_\mu
+i\,g_\pm\,A_\mu$, $g_\pm=\pm g$ is the charge of the $\phi_\pm
,\tilde\phi_\pm$ fields.  From the $N\=2$ point of view, these belong
to two hypermultiplets of opposite charge, $h_1 = (\phi_+, \phi^*_-)$
and $h_2=(\tilde\phi_-, \tilde\phi^*_+)$.  Note the {\it extra} accidental
$SU(2)$ symmetry between ($\phi_+$, $\tilde\phi_+$), and also between
($\tilde\phi_-$, $\phi_-$) due to the charge assignments. It will be
important in what follows. This $SU(2)$ is not broken when adding FI
term, whereas $SU(2)_R$ is.

The potential (\ref{pot}) has two different types of
minima:

{\bf Supersymmetric minima:} with $V=0$, that is, 
\bea & &S=0\qquad\left|\phi_+ \phi_- - \tilde\phi_+
\tilde\phi_-\right| \ = \ 0 \nonumber\\ & &|\phi_+|^2 +
|\tilde\phi_+|^2 - |\phi_-|^2 - |\tilde\phi_-|^2 \ = \ \omega^2 \ \ .
\label{v1}
\eea 

Due to the FI term, gauge symmetry is broken and the
hypersymmetric Higgs mechanism takes place; all fields acquiring mass have
$m^2=\omega^2g^2$ since we are automatically in the Bogomol'nyi
limit.

In the analogous model with only one hypermultiplet
(only $\phi_\pm$), the vacuum manifold after inflation is
simply connected and cosmic strings form via the Kibble mechanism.
In our model, the vacuum manifold is not simply connected but the
condition of finite energy per unit length still correlates the phases
of all scalars far from the string core, and causes the quantization
of magnetic flux.
Moreover, the Bogomol'nyi equations force $\phi_-=\tilde\phi_-=0$, so
there is a vacuum selection effect \cite{PRTT96,ADPU02,ADPU03}.

There are BPS string solutions involving the remaining $\phi_+,\tilde\phi_+$
fields which are SL strings \cite{VA91} 
(recall the $SU(2)$ symmetry between them).
The stability properties and cosmological formation
rates of SL strings are very different from those of NO
strings (see below).

{\bf Non-supersymmetric minima:} given by
\be
\phi_\pm=\tilde\phi_\pm=0\qquad S^2>\frac{\omega^2}{2}=S^2_c
\ee

These are local minima with potential   
$V=g^2\omega^4/8$, and $S$ is a flat direction.
Clearly, as the (false) vacuum energy is non-zero, all SUSYs
are broken. SUSY breaking causes mass splitting
within each hypermultiplet: 
\be
&m^2_\pm=S^2g^2\mp\frac{g^2}{2}\omega^2\qquad
m^2_\psi=g^2 S^2
\label{masses}
\ee where $m_\pm$ is the mass of field $\phi_\pm,\tilde\phi_\pm$,
whereas $m_\psi$ is the mass of their superpartner fermions.

\section{Inflation}
\label{inflation}

Standard D-term inflation assumes a chaotic inflationary scenario
\cite{L83}, where $S$ will have initial random values; and the regions
where $S>S_c$ will inflate. During inflation, the system is in the
false vacuum, supersymmetry is broken and the potential gets
corrections from the different masses of the bosons and fermions
(\ref{masses}).

These corrections are  known for one
hypermultiplet \cite{K01,KL03,DSS94,DM99}. Moreover, the cosmological
predictions of P-term inflationary models were investigated in
\cite{K01,KL03} and constrained by the WMAP data \cite{wmap} in
\cite{WLA03}, showing that the predictions can be
 within the observed values, but problems arise with 
the cosmic strings formed afterwards.

The correction is easily generalised to $C$ hypermultiplets ($C=2$
here) of equal $g^2$ \cite{liddle},
and in the limit $S\gg S_c$ it can be approximated by \be V_{\rm
 eff}=\frac{g^2}{8}\omega^4\left\{1+\frac{Cg^2}{8\pi^2}\left[\ln\frac{S^2g^2}{\Lambda^2}+\frac{3}{2}\right]\right\}
 \ee

This potential is a  hybrid inflation type potential \cite{liddle}. We shall follow the standard procedures to get
information about the cosmological predictions arising from this potential: 
The value of the inflaton 
when the cosmologically interesting scales leave the horizon $S_N$ can be obtained from the number $N$ of $e$-foldings:
\be
N=\frac{1}{M_{\rm Pl}}\int_{S_{\rm end}}^{S_N} \!\!\rd S\,\frac{V}{V'}
\label{n}
\ee $S_{\rm end}$ represents the end of inflation.  This will be well
before gauge symmetry breaking takes place \footnote{Inflation will
continue from the moment when slow-roll fails until the amplitude of
the oscillation$\sim S_C$. The number of $e$-foldings during this
period of inflation is typically negligible \cite{liddle}.}, because
slow-roll fails at $S\simgeq S_c$.  In any case, the precise value of
$S_{\rm end}$ is not important in evaluating $N$, since the major
contribution will come from $S_N$. In order to keep the theory under
control, we need $S_N< M_{\rm Pl}$: \be S_N\sim\sqrt{\frac{N C
g^2}{2\pi^2}}M_{\rm Pl}\leq \tau M_{\rm Pl}\Rightarrow g\leq 0.1 \ee
where we have used $N=55$ \cite{LL03}, $C=2$ and $\tau\sim 0.1$. From
the COBE normalisation, we can obtain that the energy scale $S_c$ at
which the gauge symmetry breaking will take place is similar to GUT
scale \be S_c=\frac{\omega}{\sqrt{2}}\sim
\sqrt{\frac{5.2\times10^{-4}}{2\pi}}\left(\frac{C}{N}\right)^{1/4}M_{\rm
Pl}\sim10^{16}{\rm GeV} \ee The slow-roll parameters ($\varepsilon$,
$\eta$), the spectral index ($n$) and the relative amplitude of tensor
to scalar perturbations ($R$) are within the
phenomenological values \cite{LL03b} 
\bea
\varepsilon=\frac{g^2C}{16\pi^2N}\leq2\times10^{-6}\,,\quad\eta=\frac{-1}{2N}\sim
-0.008\nonumber\\
n=1-\frac{1}{N}\sim0.98\,,\quad
R=16\varepsilon\leq3\times 10^{-5} \eea

These numbers are not very different from those in \cite{K01,KL03}, as
expected, since some parameters are independent of $C$, and in others
the difference between $C=1$ and $C=2$ is very mild.  Thus, the good
agreements with the cosmological parameters obtained for the
one-hypermultiplet case also apply here.  We now turn to the
differences, which arise after gauge symmetry breaking.

\section{Extended objects after inflation}

Once inflation is over, and the field $S$ approaches the value $S_c$,
the fields $\phi_\pm,\tilde\phi_\pm$
 will roll down to the supersymmetric vacua
(\ref{v1}). 
In the one-hypermultiplet case, both NO
cosmic strings and vortons \cite{vorton} will form via the Kibble mechanism. These defects
create perturbations to the metric of the same order of magnitude as
inflation \cite{cosmic}. This is a problem
since observational CMB data \cite{wmap} constrain the contribution of
cosmic strings to less then a few percent of the total.

By contrast, with two hypermultiplets
the defects are SL strings,
whose stability depends on the ratio of the scalar ($m_s$) and
vector ($m_v$) masses. The parameter $\beta = m_s^2/m_v^2$
(analogous to the $\kappa$ parameter that distinguishes type I
from type II superconductors) separates stable strings ($\beta <1$)
from unstable strings ($\beta >1$).

The cosmological formation and evolution of a network of such strings
was analysed in \cite{ABL99} with the conclusion that no strings will
form if $\beta \geq 1$. Here we are in the Bogomol'nyi limit, $\beta =
1$, so neither vortons nor
 strings are expected after inflation (other than perhaps a
few transient ones that will quickly disappear).  In fact, the
solution to the Bogomol'nyi equations is a {\it one-parameter family} of
magnetic vortices, degenerate in energy but with different core
structure, with string ``widths'' ranging from the NO string to wider
and wider vortices looking more and more like $CP^1$ lumps \cite{H92}. They
are all BPS states, partially breaking SUSY in their core
\cite{ADPU03}, and therefore stable. Nevertheless, there is a zero
mode linking these states 
 which, once excited, will
invariably drive the vortices to flux tubes of greater and greater
radius \cite{leese}.

The zero mode plays a very important role in preventing the
cosmological formation of the strings \cite{leese,ADPU03}, so we
should immediately worry about whether supergravity effects would
lift this degeneracy. In the bosonic case, the zero mode
survives coupling to gravity \cite{GORS}, and the same is true here:
the fattened BPS states all carry the same $U(1)$ flux and deficit
angle as the NO string, so the holonomy cancellation \cite{DKP} that is
needed for unbroken supersymmetry still holds. Thus, coupling to
supergravity will preserve all of these BPS states, and their
degeneracy, for the same reason that it preserves their topological
cousins in the minimal model \cite{DKP,AU04}. As a result, the flat
space analysis of \cite{ABL99,ADPU03} goes through and we do not
expect strings to form at the end of inflation.

Scalar gradients in this model will also contribute to the CMB
anisotropy. There are two main sources. One is the
$\phi_-,\tilde\phi_-$ fields, which have been observed to give a very
small contribution to the energy density in the one
hypermultiplet case \cite{PU03}. The other is the contribution of the
fattened vortices in the $\phi_+,\tilde\phi_+$ fields, which are
more akin to (stabilized) textures.  The CMB constraints on textures are much
weaker than those on strings \cite{durrer} but a more detailed study
is needed, in particular since textures can contribute extra power on
large angular scales.

These conclusions depend on supersymmetry not being completely broken
until much later, otherwise the results depend on how the breaking
proceeds.  If SUSY is broken due to soft mass terms, there will be
several scenarios depending on the masses \cite{PU03}: the strings may
remain SL, they can turn into NO vortices, or strings will disappear.
Also, a vorton problem may arise, depending on the nature of the
breaking and the detailed dynamics of the zero mode, if the latter
survives \cite{ADPU03,vorton}.

For completeness, we give an estimate of the CMB signal deep in the
stability region $\beta < 1$ assuming that the $SU(2)$ symmetry is not
broken. In this case SL strings can form with a density comparable to
NO strings, and scaling behaviour is expected, but the reduced mass
per unit length also weakens the CMB signal \footnote{It should be
stressed that the CMB constraints are based on numerical simulations
of NO strings that are all in the Bogomol'nyi limit. The mass per unit
length (and therefore the deficit angle and the CMB signal) of the
strings goes down with lower $\beta$.  }. The net result should be a
contribution to the CMB anisotropy which is a few percent (2-5\%) of
that of NO strings, and fairly insensitive to $\beta$.

Finally, the connection with superstrings leaves open a number of
interesting questions. First, the FI parameters are expected to be
spacetime dependent, and such cosmological models have not been
investigated to our knowledge.  Second, if several vector multiplets
are present (as it is the case in most brane and conifold models)
inflation could be driven by multiple scalar fields, and the situation
could be more complicated \cite{LR99}.

\acknowledgments

We thank N. Bartolo, E. Bergshoeff, B. de Carlos, M. Esole, F. Freire,
M. Kunz,  A.R. Liddle, A. van Proeyen and S. Vandoren for useful discussions.
 J.U. is supported by the Spanish {\it Secretar\'\i a de Estado de
 Educaci\'on y Universidades} and {\it Fondo Social Europeo}. This
 work was also partially supported by PPARC, ESF COSLAB programme, FPA
 2002-02037 and 9/UPV00172.310-14497/2002.

\end{document}